\documentclass[12pt,preprint]{aastex}

\begin{document}

\title{The Unusual Spectrum of Comet 96P/Machholz}
\author{Laura E. Langland-Shula and Graeme H. Smith}
\affil{Department of Astronomy and Astrophysics, University of California, Santa Cruz \\ 1156 High Street, Santa Cruz, CA 95064}
\email{laura@ucolick.org, graeme@ucolick.org}

\shorttitle{The Spectrum of 96P/Machholz}
\shortauthors{Langland-Shula \& Smith}

\begin{abstract}
We report spectra from 3000-5900 \AA\ for comet 96P/Machholz, obtained on 2007 Apr 27 UT with the 3-m Shane telescope at Lick Observatory. The spectra are extremely carbon poor, and show a prominent NH$_2$ series, but no CN emission. NH, NH$_2$, and C$_2$ gas production rates are (8.36 $\pm$ 2.18)$\times$10$^{25}$, (29.88 $\pm$ 3.66)$\times$10$^{25}$, and (4.52 $\pm$ 0.61)$\times$10$^{23}$ molecules sec$^{-1}$, respectively, as determined from Haser model fits to the data. Upper limits to the gas production rates for CN and C$_3$ are 7.5$\times$10$^{22}$ and 2.0$\times$10$^{23}$, respectively. Though 96P is depleted in C$_2$ and C$_3$ relative to NH, it is even more depleted in CN than other so-called ``carbon-chain depleted'' comets.
\end{abstract}


\section{INTRODUCTION}

In recent decades a rapidly growing number of periodic comets have been discovered, including new members of the low-inclination short-period Jupiter family, and the high-inclination medium-period Halley family. A sub-class of Jupiter family comets has been shown by many authors \citep{NS84,A+95,FH96} to be depleted in the ``carbon-chain'' molecules C$_2$ and C$_3$ relative to CN. A few comets with more abnormal production rate ratios have been reported, including notably C/1998 Yanaka (1988r; Fink 1992). \citet{S07} (hereafter S07) recently reported unusual gas production rate ratios for 96P, determined from narrowband photometry performed on 2007 May 12 and 24 when the comet was at $r_h$ = 1.07 and 1.30 AU, which we now confirm.

Comet 96P/Machholz was discovered in May 1986 \citep{M+86}. Surprisingly short-period (5.2 years) orbital elements were calculated from the initial magnitudes and positions reported in the IAU circulars (Table \ref{96P}). \citet{S90a} noted a major outburst several weeks after perihelion and suggested that the comet remained undiscovered in spite of its short period because it is inactive for much of the time. Its small perihelion (0.12 AU) leads to a high erosion rate \citep{S90b}, so it may have a mostly dormant iceless surface. Comet 96P was again reported to the IAU on its next apparition in 1991, reported at about 3 AU from the Sun in 1997, and observed by SOHO on three occasions \citep{M+97,L+02,G+04}.

96P is dynamically interesting. It is the shortest-period Halley family comet known. \citet{G+90} noted that 96P (then known as comet 1986VIII) travels closer to the Sun than all other known comets with orbital periods less than 150 years. It is in a 9/4 resonance with Jupiter, and their orbit evolution calculations show that its perihelion is steadily decreasing to a minimum of 0.03 AU in about the year 2450. 96P has also been suggested as the parent body for several meteor streams \citep{S90a,M90,BO93}. All other references to 96P discuss its meteor stream connection \citep{J+97,WC98}, and the possibility that some sunskirting comets are fragments of it that were thrust onto their current parabolic orbits after encounters with Jupiter \citep{OY03,SC05}.

\section{OBSERVATIONS}

Comet 96P/Machholz was observed post-perihelion on 2007 April 27 UT with the Kast Spectrometer at the Lick Observatory 3-m Shane telescope. The blue side was used with a 1.5$\times$128 arcsec slit and a 452/3306 grism giving a dispersion of 2.49 \AA/pixel. A Reticon CCD recorded the 2-D spectra (1199 pixels along the dispersion direction and 164 pixels along the slit at a scale of 0.78 arcsec/pixel). The comet's orbital parameters and observational circumstances are given in Table \ref{96P}.

The data reduction used standard IRAF\footnote{http://iraf.noao.edu/} techniques and custom IDL routines. In IRAF, the spectra were trimmed to encompass the length of the slit, flat-field corrected, median filtered to remove cosmic rays, and wavelength calibrated using HeHgCd lamp reference spectra.

Because 96P rose shortly before morning astronomical twilight, there was time for one dark-sky comet frame to be taken starting at 11:38 UT, but not enough time to obtain a separate sky spectrum for background subtraction. The second comet spectrum was taken mostly between astronomical and nautical twilight and contained a large scattered sky continuum component. Fortunately, both 2-D spectra contain sky information at the ends of the slit, with only a small amount of flux contribution from the NH band at 3360 \AA. In IDL, we constructed a virtual 2-D sky frame for each comet exposure by fitting a line under the NH flux at both the top and bottom of the slit to remove it, and linearly interpolating the sky values from the top and bottom rows of the CCD frame across the entire slit length. Back in IRAF, each virtual sky frame was subtracted from its comet frame.

\begin{figure}
\plotone{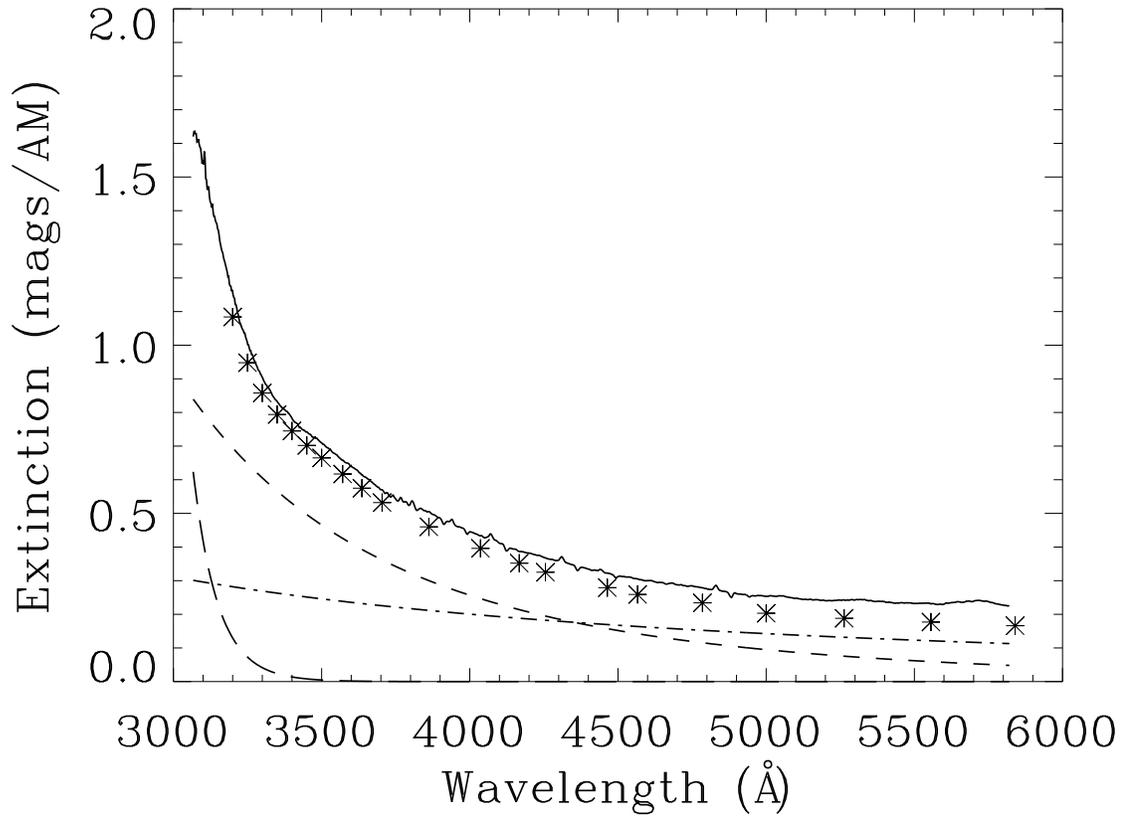}
\caption{Atmospheric extinction in magnitudes per airmass versus wavelength at Lick Observatory on 2007 Apr 27 UT. The solid line is the measured extinction. The long-dashed line is the ozone component. The short-dashed line is the Rayleigh scattering component. The dot-dashed line is the aerosol component. The crosses show extinction values tabulated previously for Lick Observatory.}
\label{ExtFig}
\end{figure}

The final flux calibration was done with a custom IDL code that applies an extinction curve measured for that night, and utilizes standard fluxes blueward of 3200\AA\ not available with IRAF. The spectrograph range was set this blue to observe the OH 3085\AA\ band in other target comets. The extinction curve was constructed from observations of the standard star BD+33 2642 over the airmass range 2.52-1.12, using a wide (9 arcsec) slit to capture all of the starlight. The extinction versus wavelength relation that was derived is shown in Fig. \ref{ExtFig}.

The IDL flux-calibration code does several things: integrates a spectrum of the standard star (taken with the same 1.5 arcsec slit as the comet) along the slit, corrects the integrated standard star spectrum to zero airmass, and creates a flux-per-count ``sensitivity function'' by dividing the standard fluxes by the integrated standard star counts for each pixel along the dispersion direction. The standard fluxes were taken from two sources: the magnitudes per unit frequency distributed with IRAF redward of 3200 \AA, and the flux per unit wavelength from \citet{B96} blueward of 3200\AA. This ``sensitivity function'' takes into account total system throughput, including wavelength-dependent light losses from using a narrower slit. The last IDL task is to calculate the appropriate extinction correction for each comet frame, and fold the comet extinction correction into the sensitivity function to create a set of final flux-calibration masks. Finally, back in IRAF, each 2-D comet spectrum is multiplied by its flux-calibration mask.

\begin{figure}
\plotone{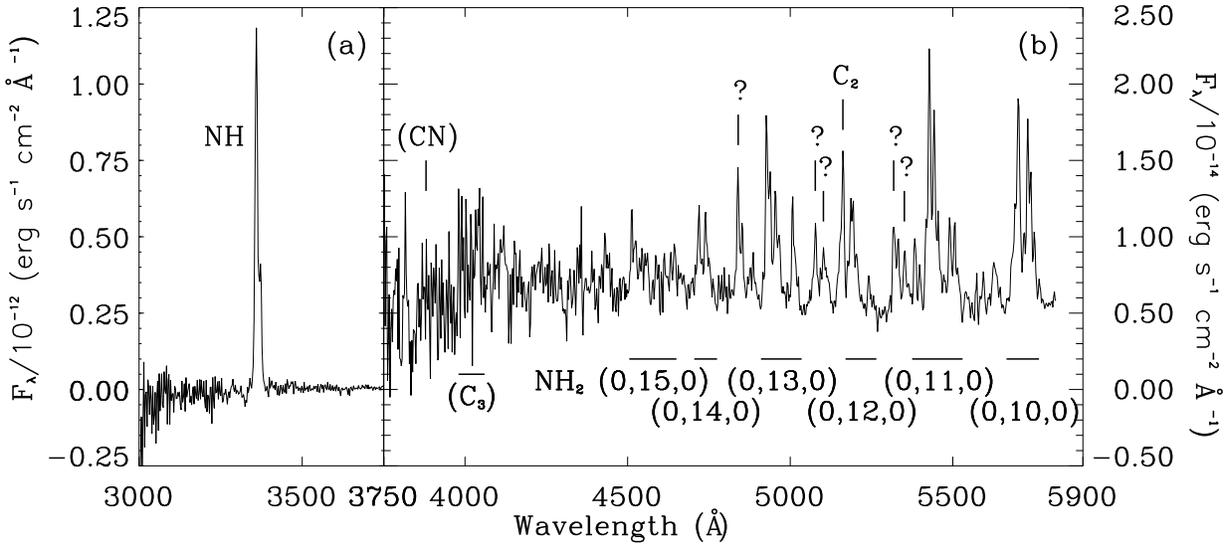}
\caption{Average integrated spectrum of 96P. (a) Flux per unit wavelength, scaled to 10$^{-12}$ erg s$^{-1}$ cm$^{-2}$ \AA$^{-1}$. Note the strong NH line at 3360 \AA. (b) The same averaged spectrum, scaled to 10$^{-14}$ erg s$^{-1}$ cm$^{-2}$ \AA$^{-1}$, to clearly show the NH$_2$ series from 4500-5800 \AA\ and the faint C$_2$ band at 5165 \AA. The CN band and the C$_3$ band at 4050 \AA\ are labeled in parentheses to denote upper limits. The question marks refer to lines that are unidentified in the high-resolution comet spectral atlases of \citet{B+96} and \citet{CC02}.}
\label{SpectFig}
\end{figure}

The average of our two flux-calibrated 1200-sec spectra of comet 96P, integrated along the slit, is shown in Fig.~\ref{SpectFig}. The NH emission band at 3360 \AA\ is clearly visible in Fig.~\ref{SpectFig}a. Various NH$_2$ emission bands are identified in Fig.~\ref{SpectFig}b. Most of the identified lines are NH$_2$ and NH, presumed to be photodissociation daughter species of ammonia. Notable is the faint C$_2$ emission, and the questionable CN and C$_3$ emission. Some flux was apparent in the region of the C$_3$ band for one spectrum, but the other spectrum only showed noise in this location.

\section{MOLECULE PRODUCTION RATES}

Molecule production rates were calculated from a \citet{H57} model. For a detailed explanation of how this model was applied to KAST data, please see \citet{TS96}. Fluorescence efficiencies are from \citet{K+89} for NH, \citet{NS89} for C$_2$ and C$_3$, and \citet{T84} for CN. For NH$_2$, we used the blue band fluorescence efficiencies of \citet{TW89}, scaled to the newer red band efficiencies of \citet{KW02}. Following \citet{A+95}, the gas outflow velocity scales with heliocentric distance as $1.0 \times r_h({\rm AU})^{-0.5}$ km s$^{-1}$. Sets of scale lengths were determined for the NH, NH$_2$, and C$_2$ bands in each spectrum by experimenting with a range of possible values. Where the best-fit scale lengths for NH$_2$ bands blueward of (0,12,0) noisily exceeded 100$\times$10$^3$ km, the mean of the other cleaner scale lengths was used instead. Scale lengths for the CN and C$_3$ bands followed from \citet{TS96}.

The scale lengths (applicable at the comet's heliocentric distance $r_h = 0.754$ AU), and the resulting gas production rates ($Q$) with 1-$\sigma$ error bars, are given in Table \ref{ProdRates}. ``NA'' in Table \ref{ProdRates} refers to an undefined production rate where the net flux in a non-detected emission band was less than the adjacent continuum. S07 report log $Q$(OH) = 27.33, while we did not detect OH. Our production rate for NH is a factor of 3 higher than S07, consistent with our smaller $r_h$ than S07, but our production rate for C$_2$ is 90\% of S07's. Our upper limits for CN and C$_3$ are consistent with S07. 96P may be sputtering somewhat unevenly as its global gas production turns off.

\section{DISCUSSION}

The only other comet in the literature noted to be this deficient in CN and C$_2$ is Yanaka (1988r or C/1998 Y1; Kosai et al. 1988). Yanaka was observed post-perihelion in 1989 with the f/1.2 spectrograph at the University of Arizona 1.54-m Mount Bigelow telescope \citep{F91,F92}. The observational parameters are listed in Table \ref{C1998Y1}; where not given in \citet{F92}, they were looked up at the Minor Planet \& Comet Ephemeris Service\footnote{http://www.cfa.harvard.edu/iau/MPEph/MPEph.html} provided by the IAU.

Comet Yanaka displays a series of NH$_2$ emission bands from 5300-8500\AA. Fink's spectrum of it and our spectrum of 96P/Machholz overlap from 5300-5800\AA, in the region of the NH$_2$ (0,10,0) and (0,11,0) bands. Noticeably absent in the Yanaka spectrum are the CN 1-0, 2-0, 2-1, and 3-1 bands, and the prominent C$_2$ $\Delta {\rm v} = -1$ band.

\citet{G+93} proposed that three effects could combine non-linearly to inhibit the detection of CN and C$_2$ in Comet Yanaka. First, cosmic ray exposure could ``glue together'' or ``carbonize'' the comet (deplete H, N, and O relative to C), with two effects: trapping carbon-bearing gas molecules between dust grains, and producing larger dust grains more resistant to heating and fragmentation. Lastly, since Comet Yanaka was observed over a smaller spatial dimension (660 $\times$ 16,470 km) than the other comets in Fink (1992; 3420 $\times$ 84,200 km and 2400 $\times$ 145,600 km), there may be less time for molecule production to occur within the spectrograph slit aperture.

Our observation of 96P covers a spatial area of about 820 $\times$ 70,570 km. Our slit length is comparable to the larger spatial observations in \citet{F92}. \citet{C85} reviews Haser-model decay scale lengths for the parent molecules of CN (1.2-2.2$\times$10$^4$ km), C$_2$ (1.7-3.5$\times$10$^4$ km), and C$_3$ (1.0-2.5$\times$10$^3$ km), which are all shorter than our slit length. If cosmic ray gluing increased the parent decay scale length of C$_3$ by an order of magnitude, we still expect to see C$_3$ production. Brightness and polarization measurements of dust in the coma of 96P \citep{G+04} show grain sizes consistent with in-situ measurements of Halley dust, also pointing to limited cosmic ray gluing.
 
\citet{G+93} admit that other dynamically new comets do not display extreme carbon depletions. Likewise, other short-period comets ($P <$ 7 years), presumably with surfaces similarly processed by solar radiation as 96P, display measurable carbon production: 2P/Encke and 6P/d'Arrest \citep{A+79}, 9P/Tempel 1 \citep{L+06}, 19P/Borrelly \citep{H+02}, 26P/Grigg-Skjellerup \citep{J+93}, 46P/Wirtanen \citep{J+98}, and 67P/Churyumov-Gerasimenko \citep{S+04}. In old ``dusty'' comets, or comets moving outbound from perihelion, the CN and C$_2$ emission bands are the last to be observable as global gas production turns off (i.e. A'Hearn et al. 1995, Fink \& Hicks 1996), such that CN is used as a gas-to-dust indicator (i.e. Storrs et al. 1992). However, CN is not detected in the Kast spectra of 96P.

Other dormant comets may perhaps show mostly ices, and not carbon-bearing gas, when they outburst. However, CN was detected in 95P/Chiron by \citet{B+91}. Comet 4P/Faye was observed in outburst by \citet{G96}, and has measurable CN, C$_2$, and C$_3$ production rates \citep{GL94}. \citet{C+96} detected no CN in three candidate comet-asteroid transition objects, although their signal-to-noise deteriorates in the region of the CN band. Unfortunately, none of these observations were programmed to look for the NH 3360 \AA\ band.

Having exhausted several explanations for the odd spectrum of 96P, we can compare it to other comets with extreme production-rate ratios. The values in square brackets in Table \ref{Extreme} refer to logarithmic ratios between the production rates of various species. Where there are discrepancies between the [X/OH] and the [X/CN] ratios for the \citet{A+95} comets, the [X/CN] ratios are more reliable, and are the basis for the listed [X/NH] ratios. In 96P, we confirm the S07 report that the carbon-chain molecules are depleted compared to NH. Even compared to other ``C-chain-depleted'' comets 96P is unusual because we detect C$_2$ emission but not CN. A distinctive property of 96P is the very high [C$_2$/CN] ratio, for which we only obtain a lower limit. Rather than being carbon-chain depleted relative to CN, 96P is the comet with the least amount of CN to also display a C$_2$ band, and by far the comet most depleted in CN relative to NH, C$_2$, and C$_3$ than either ``typical'' or ``C-chain-depleted'' comets.

\section{CONCLUSIONS}

Comet 96P/Machholz has extremely depleted C$_2$ and C$_3$ relative to NH and NH$_2$, with CN more extremely depleted. The dimensions of our spectrograph slit were sufficient to see any possible decay of the parents of these carbon molecules. The extreme carbon depletion of 96P is unlikely to be due to cosmic ray ``gluing'' in the comet's past, and unlikely to be due to surface processing by the Sun over repeated short orbits (since this effect is not seen in many other short-period comets). Though 96P was observed to have an outburst shortly after its discovery, other confirmed outbursting comets show carbon emission features. Thus, it appears that 96P/Machholz belongs to a small class of comets with genuinely unusual molecule production rates.

\acknowledgments
This research is supported by NASA grant NNG05GG59G through the Planetary Astronomy program.


\clearpage

\begin{deluxetable}{llllll}
\tablecaption{Comet 96P/Machholz: Observations and Orbit Parameters\label{96P}}
\tablehead{
\multicolumn{2}{c}{UT} & \colhead{$r_h$ (AU)} & \colhead{$\Delta$ (AU)} & \colhead{Airmass} & \colhead{Exp (s)}
}
\startdata
\multicolumn{2}{l}{2007 Apr 27 11:38} & 0.754 & 0.747 & 2.5-1.87 & 2 $\times$ 1200 \\
\hline
\multicolumn{2}{l}{$q = 0.1241 AU$} & \multicolumn{2}{l}{$e = 0.9588$} & \multicolumn{2}{l}{$i = 60.1813^\circ$} \\
\enddata
\end{deluxetable}

\clearpage

\begin{deluxetable}{lrrrrrrr}
\tabletypesize{\footnotesize}
\tablecaption{Haser Model Scale Lengths and Gas Production Rates $Q$ for Comet 96P/Machholz\label{ProdRates}}
\tablehead{
\colhead{} & \multicolumn{3}{c}{Spectrum 1} & \multicolumn{3}{c}{Spectrum 2} & \colhead{Mean} \\
\colhead{Molecule} & \colhead{Parent} & \colhead{Daughter} & \colhead{$Q$} & \colhead{Parent} & \colhead{Daughter} & \colhead{$Q$} & \colhead{$Q$} \\
\colhead{} & \colhead{(10$^3$ km)} & \colhead{(10$^3$ km)} & \colhead{(10$^{25}$ mol. s$^{-1}$)} & \colhead{(10$^3$ km)} & \colhead{(10$^3$ km)} & \colhead{(10$^{25}$ mol. s$^{-1}$)} & \colhead{(10$^{25}$ mol. s$^{-1}$)}
}
\startdata
NH & 70 & 200 & 10.54 & 40 & 120 & 6.18 & 8.36 $\pm$ 2.18 \\
NH$_2$ mean & 3.4 & 42 & 26.22 & 4.2 & 25 & 33.54 & 29.88 $\pm$ 3.66 \\
-(0,10,0) & 3.0 & 50 & 17.07 & 6.0 & 25 & 23.25 & 20.16 $\pm$ 3.09 \\
-(0,11,0) & 3.0 & 40 & 20.3 & 4.0 & 25 & 21.65 & 20.98 $\pm$ 0.68 \\
-(0,12,0) & 5.0 & 35 & 13.03 & 5.0 & $>$300 & 29.15 & 21.09 $\pm$ 8.06 \\
-(0,13,0) & 2.0 & 160 & 47.14 & 3.0 & 150 & 64.47 & 55.81 $\pm$ 8.67 \\
-(0,14,0) & 4.0 & $>$300 & 33.56 & 3.0 & 210 & 29.19 & 31.38 $\pm$ 2.18 \\
CN & 7.5 & 240 & NA & 7.5 & 240 & $<$0.0075 & $<$0.0075 \\
C$_2$ & 7.0 & 20 & 0.039 & 7.0 & 30 & 0.051 & 0.0452 $\pm$ 0.0061 \\
C$_3$ & 1.3 & 57 & $<$0.020 & 1.3 & 57 & NA & $<$0.020 \\
\enddata
\end{deluxetable}

\clearpage

\begin{deluxetable}{llllll}
\tablecaption{C/1998 Y1: Observations (Fink 1992) and Orbit Parameters\label{C1998Y1}}
\tablehead{
\multicolumn{2}{c}{UT} & \colhead{$r_h$ (AU)} & \colhead{$\Delta$ (AU)}  & \colhead{Airmass} & \colhead{Exp (s)}
}
\startdata
\multicolumn{2}{l}{1989 Jan 15 13:26} & 0.932 & 0.367 & 1.84 & 3 $\times$ 300 \\
\hline
\multicolumn{2}{l}{$q = 0.4278 AU$} & \multicolumn{2}{l}{$e = 1.0000$} & \multicolumn{2}{l}{$i = 71.0077^\circ$} \\
\enddata
\end{deluxetable}

\clearpage

\begin{deluxetable}{llllllllll}
\tabletypesize{\scriptsize}
\tablecaption{Comets With Extreme Production Rate Ratios\label{Extreme}}
\tablehead{
\colhead{Comet} & \colhead{Type} & \colhead{[CN/OH]} & \colhead{[C$_2$/OH]} & \colhead{[NH/OH]} & \colhead{[C$_2$/CN]} & \colhead{[CN/NH]} & \colhead{[C$_2$/NH]} & \colhead{[C$_3$/NH]} & \colhead{Ref.}
}
\startdata
``Typical'' mean & T & -2.50 & -2.44 & -2.37 & +0.06 & -0.13 & -0.07 & -1.22 & 1 \\
``C-Chain Dep.'' mean & D & -2.69 & -3.30 & -2.48 & -0.61 & -0.21 & -0.82 & -1.70 & 1 \\
P/Howell & JF,T & -2.47 & -2.55 & \textbf{-1.14} & -0.09 & \textbf{-1.33} & \textbf{-1.42} & \textbf{-2.68} & 1 \\
Bowell (1980b) & DN,D & \textbf{-3.33} & -3.43 & -2.06 & -0.39 & \textbf{-0.58} & -0.97 & -2.26 & 1 \\
P/IRAS & HF,D & -2.57 & -3.48 & -2.03 & -0.93 & -0.45 & \textbf{-1.38} & -2.28 & 1 \\
P/Wolf-Harrington & JF,D & -2.53 & \textbf{-3.75} & \textbf{-2.78} & \textbf{-1.20} & \textbf{+0.30} & -0.90 & -1.75 & 1 \\
Yanaka (1988r) & DN,D & \textbf{$<$-3.97} & \textbf{$<$-4.34} & \textbf{-3.14} & ... & \textbf{$<$-0.83} & \textbf{$<$-1.20} & ... & 2 \\
96P/Machholz & HF,E & \textbf{-4.93} & \textbf{-3.63} & \textbf{-1.86} & \textbf{+1.30} & \textbf{-3.07} & \textbf{-1.77} & \textbf{-3.17} & 3\\
96P (this work) & HF,E & ... & ... & ... & \textbf{$>$+0.78} & \textbf{$<$-3.04} & \textbf{-2.26} & \textbf{$<$-2.61} & ...\\
\enddata
\tablecomments{Comet families: JF=Jupiter, HF=Halley, DN=Dynamically New. ``Carbon-chain depletion'' relative to CN: T=Typical, D=Depleted, E=Enhanced. Very unusual ratios are highlighted in \textbf{bold-face}.}
\tablerefs{(1) A'Hearn et al. 1995; (2) Fink 1992; (3) Schleicher 2007}
\end{deluxetable}

\clearpage

\clearpage

\begin{thebibliography}{}
\bibitem[A'Hearn, Millis, \& Birch (1979)]{A+79} A'Hearn, M. F., Millis, R. L., and Birch, P. V. 1979, AJ, 84, 570
\bibitem[A'Hearn et al. (1995)]{A+95} A'Hearn, M. F., Millis, R. L., Schleicher, D. G., Osip, D. J., \& Birch, P. V. 1995, Icar, 118, 223
\bibitem[Babadzhanov \& Obrubov (1993)]{BO93} Babadzhanov, B. P. \& Obrubov, V. Yu. 1993, mtpb.conf, 49
\bibitem[Bohlin (1986)]{B96} Bohlin, R. C. 1986, ApJ, 308, 1001
\bibitem[Brown et al. (1996)]{B+96}Brown, M. E., Bouchez, A. H., Spinrad, A. H., \& Johns-Krull, C. M. 1996, AJ, 112, 1197
\bibitem[Bus et al. (1991)]{B+91}Bus, S. J., A'Hearn, M. F., Schleicher, D. G., \& Bowell, E. 1991, Sci, 251, 774
\bibitem[Chamberlin et al. (1996)]{C+96}Chamberlin, A. B., McFadden, L., Schulz, R., Schleicher, D. G., \& Bus, S. J. 1996, Icar, 119, 173
\bibitem[Cochran (1985)]{C85} Cochran, A. 1985, AJ, 90, 2609
\bibitem[Cochran \& Cochran (2002)]{CC02}Cochran, A. L., \& Cochran, W. D. 2002, Icar, 157, 297
\bibitem[Fink (1991)]{F91} Fink, U. 1991, BAAS, 23, 1160
\bibitem[Fink (1992)]{F92} Fink, U. 1992, Sci, 257, 1926
\bibitem[Fink \& Hicks (1996)]{FH96}Fink, U. \& Hicks, M. D. 1996, ApJ, 459, 729
\bibitem[Gil-Hutton \& Licandro (1994)]{GL94}Gil-Hutton, R. \& Licandro, J. 1994, RMxAA, 28, 3
\bibitem[Green et al. (1990)]{G+90}Green, D. W. E., Rickman, H., Porter, A. C., \& Meech, K. J. 1990, Sci, 247, 1063
\bibitem[Greenberg, Singh, \& de Almeida (1993)]{G+93} Greenberg, J. M., Singh, P. D., \& de Almeida 1993, ApJ, 414, L45
\bibitem[Grothues (1996)]{G96}Grothues, H. 1996, P\&SS, 44, 625
\bibitem[Grynko, Jockers, \& Schwenn (2004)]{G+04} Grynko, Y., Jockers, K., \& Schwenn, R. 2004, A\&A, 427, 755
\bibitem[Hamane et al. (2002)]{H+02} Hamane, T., Kawakita, H., Kinugasas, K., Yamamura, T., \& Takeyama, N. 2002, PASJ, 54, L35
\bibitem[Haser (1957)]{H57} Haser, T. 1957, BSRSL, 43, 740
\bibitem[Jenniskens et al. (1997)]{J+97} Jenniskens, P., Betlem, H., de Lignie, M., Langbroek, M., \& van Vliet, M. 1997, A\&A, 327, 1242
\bibitem[Jockers et al. (1993)]{J+93} Jockers, K., Kiselev, N. N., Boehnhardt, H., \& Thomas, N. 1993, A\&A, 268, L9
\bibitem[Jockers, Credner, \& Bonev (1998)]{J+98} Jockers, K., Credner, T., \& Bonev, T. 1998, A\&A, 335, L56
\bibitem[Kawakita \& Watanabe (2002)]{KW02}Kawakita, H. \& Watanabe, J. 2002, ApJ, 572, L177
\bibitem[Kim, A'Hearn, \& Cochran (1989)]{K+89}Kim, S. J., A'Hearn, M. F., \& Cochran, W. D. 1989, Icar, 77, 98
\bibitem[Kosai, Yanaka, \& Levy (1988)]{K+88} Kosai, H., Yanaka, T., \& Levy, D. 1988, IAUC, 4696, 1
\bibitem[Lara et al. (2006)]{L+06} Lara, L. M., Boehnhardt, H., Gredel, R., Guti\'{e}rrez, P. J., Ortiz, J. L., Rodrigo, R., \& Vidal-Nu\~{n}ez, M. J. 2006, A\&A, 445, 1151
\bibitem[Lisse, Fernandez, \& Biesecker (2002)]{L+02} Lisse, C. M., Fernandez, Y. R., \& Biesecker, D. A. 2002, DPS, 34, 1206
\bibitem[Machholz, Morris, \& Hale (1986)]{M+86} Machholz, D. E., Morris, C. S., \& Hale, A. 1986, IAUC, 4214, 1
\bibitem[McIntosh (1990)]{M90} McIntosh, B. A. 1990, Icar, 86, 299
\bibitem[Meech et al. (1997)]{M+97} Meech, K. J., Hainaut, O. R., Bauer, J. M., Williams, G. V., St. Cyr, O. C., \& Stezelberger, S. T. 1997, IAUC, 6669, 2\
\bibitem[Newburn \& Spinrad (1984)]{NS84}Newburn, R. L. \& Spinrad, H. 1984, AJ, 89, 289
\bibitem[Newburn \& Spinrad (1989)]{NS89}Newburn, R. L. \& Spinrad, H. 1989, AJ, 97, 552
\bibitem[Ohtsuka, Nakano, \& Yoshikawa (2003)]{OY03} Ohtsuka, K., Nakano, S., \& Yoshikawa, M. 2003, PASJ, 55, 321
\bibitem[Schleicher (2007)]{S07}Schleicher, D. 2007, IAUC, 8842 (S07)
\bibitem[Schulz, St\"{u}we, \& Boehnhardt (2004)]{S+04} Schulz, R., St\"{u}we, J. A., \& Boehnhardt, H. 2004, A\&A, 422, L19
\bibitem[Sekanina (1990a)]{S90a} Sekanina, Z. 1990, AJ, 99, 1268 (1990a)
\bibitem[Sekanina (1990b)]{S90b} Sekanina, Z. 1990, AJ, 100, 1293 (1990b)
\bibitem[Sekanina \& Chodas (2005)]{SC05} Sekanina, Z. \& Chodas, P. W. 2005, ApJS, 161, 551
\bibitem[Storrs, Cochran, \& Barker (1992)]{S+92} Storrs, A. D., Cochran, A. L., \& Barker, E. S. 1992, Icar, 98, 163
\bibitem[Tatum (1984)]{T84}Tatum, J. B. 1984, A\&A, 135, 183
\bibitem[Tegler \& Wyckoff (1989)]{TW89}Tegler, S. \& Wyckoff, S. 1989, ApJ, 343, 445
\bibitem[Turner \& Smith (1996)]{TS96}Turner, N. J. \& Smith, G. H. 1996, EM\&P, 73, 33
\bibitem[Williams \& Collander-Brown (1998)]{WC98} Williams, I. P. \& Collander-Brown, S. J. 1998, MNRAS, 294, 127
\end{thebibliography}
\end{document}